\documentstyle[preprint,aps]{revtex} 
\begin{document}
\draft
\date{to appear in {\sl The Philosophical Magazine B}}
\title{Optical absorption of strongly 
correlated half-filled Mott-Hubbard chains}
\author{F.~Gebhard\footnote{e-mail: {\tt florian\verb2@2gaston.ill.fr}}}
\address{ILL Grenoble, B.~P.\ 156x, F-38042 Grenoble Cedex 9, France}
\author{K.~Bott, M.~Scheidler, P.~Thomas, S.~W.~Koch}
\address{Dept.~of Physics and Materials Sciences Center,\\
Philipps University Marburg, D-35032~Marburg, Germany}

\maketitle

\begin{abstract}%
In this last of three articles on the optical absorption
of electrons in a half-filled Peierls-distorted chain
we address the dimerized extended Hubbard model in the limit
of a large on-site interaction~$U$. When the Hubbard interaction
is large both compared to the band width~$W$ and the 
nearest-neighbor interaction~$V$ the 
charge dynamics is properly described by the
Harris-Lange model. This model can be exactly mapped onto
a model of free spinless Fermions in parallel (Hubbard-)bands of width~$W$
which are eventually Peierls-split.

To determine the coherent absorption features at low temperatures
we design and employ the ``no-recoil approximation'' in which we
assume that the momentum transfer to the spin degrees of freedom
can only be $\Delta q_S=0$ or $\Delta q_S=\pi/a$
during an optical excitation. 
We present explicit analytical results for the optical
absorption in the presence
of a lattice dimerization~$\delta$ and a nearest-neighbor interaction~$V$
for the N\'{e}el and dimer state.
We find that the coherent part of the optical absorption for $V=0$
is given by a single peak at~$\omega=U$ and broad but weak
absorption bands for $W\delta\leq |\omega-U| \leq W$. 
The central peak at $\omega=U$ only vanishes  for $\delta=0$
in the N\'{e}el state.
For an appreciable
nearest-neighbor interaction~$V>W/2$ almost all spectral weight
is transferred to the $\Delta q_C=0$-exciton and the 
$\Delta q_C=\pi/a$-exciton
whose relative spectral weights very sensitively depend on {\em both\/}
the lattice {\em and\/}
the spin dimerization of the ground state.

\vskip2cm\noindent
PACS1996: 71.10.Fd, 71.20.Rv, 36.20.Kd 
\end{abstract}

\newpage
\section{Introduction}

In two previous articles
(Gebhard, Bott, Scheidler, Thomas, and Koch I and II 1996)     
we have studied the optical absorption of non-interacting and
strongly interacting tight-binding
electrons in Peierls-distorted chains.
These two different cases are idealizations of
real almost ideal one-dimensional semiconductors (Farges 1994).    
While polymers like polyacetylen
(Heeger, Kivelson, Schrieffer, and Su 1988);                       
(Baeriswyl, Campbell, and Mazumdar 1992); (Schott and Nechtschein 1994)
are commonly believed to be Peierls insulators
some charge-transfer salts (Alc\'{a}cer 1994); (Brau and Farges 1994)   
should be understood as Mott-Hubbard insulators.
The extended Hubbard model for interacting electrons
on a distorted chain at half-filling is considered appropriate for the latter
class of materials (Mazumdar and Dixit 1986);
(Fritsch and Ducasse 1991); (Mila 1995).                            
Unfortunately, the study of the optical, i.e., finite
frequency properties of correlated electron systems
poses a very difficult many-body problem
that cannot be solved analytically without further approximations
on the Hamiltonian or the calculations
(Kohn 1964); (Maldague 1977); (Lyo and Galinar 1977); (Lyo 1978);
(Galinar 1979);    
(Campbell, Gammel, and Loh 1989); (Mahan 1990); (Shastry and Sutherland 1990);
(Stafford, Millis, and Shastry 1991); 
(Fye, Martins, Scalapino, Wagner, and Hanke 1992); (Stafford and Millis 1993).

In our second article (Gebhard {\em et al.} II 1996)       
we focused on the optical
absorption in the dimerized extended Harris-Lange model 
which is equivalent to the Hubbard model to lowest order
in the strong-coupling expansion. We could exactly map
the charge degrees of freedom
onto two parallel (Hubbard-)bands 
for free spinless Fermions of band-width~$W$.
The eigenstates of the Harris-Lange model are highly spin-degenerate 
which allowed us to exactly calculate the optical absorption.
The results apply to the ``hot-spin'' regime of the Hubbard model
when the temperature is large compared to the spin exchange 
energy~$J\sim W^2/(U-V)$. Real experiments are carried out at low temperatures
for which the system
is in an unique ground state with antiferromagnetic correlations. 
Unfortunately, this problem cannot be solved analytically.
In this article we will
employ the analogy to an ordinary semiconductor (electrons and holes
in a phonon bath) to design a ``no-recoil approximation''
for the ``chargeons'' in a ``spinon bath''.
It will allow us to determine the coherent absorption features 
of the Hubbard model at large~$U/W$.

The paper is organized as follows.
In section~\ref{Hamilts} we derive our approximate Hamiltonian from 
the Hubbard model. The charge dynamics will be governed
by the Harris-Lange model while the ground state is determined
by the Heisenberg model.
Basic results of our articles (Gebhard {\em et al.} I and II 1996)   
for the optical absorption and the current operator are recalled
in section~\ref{optabssec}.
In section~\ref{norecoilsec} we introduce the ``no-recoil approximation'' 
which allows us to obtain informations on the coherent part of
the absorption spectrum at low temperatures. The corresponding results
for the Peierls-distorted extended Harris-Lange model are presented in
section~\ref{optabsHubbard}.
Summary and conclusions close our presentation.

\section{Model Hamiltonians}
\label{Hamilts}

\subsection{Charge degrees of freedom at strong coupling: 
the Harris-Lange model}

As shown in (Gebhard {\em et al.} II 1996)                           
the Peierls-distorted extended 
Hubbard model (Hubbard 1963); (E\ss ler and Korepin 1994)        
can be mapped onto the Harris-Lange
model in the limit of strong correlations.
In standard notation of second quantization the latter model reads 
\begin{mathletters}
\begin{eqnarray}
\hat{H}_{\rm HL}^{\rm dim, \, ext} &=& \hat{T}_{\rm LHB}(\delta) +
 \hat{T}_{\rm UHB}(\delta) + U\hat{D} + V \hat{V}\\[6pt]
\hat{T}_{\rm LHB} &=& (-t) \sum_{l,\sigma} (1+(-1)^l\delta) 
\left(1-\hat{n}_{l,-\sigma}\right)
\left(
\hat{c}_{l,\sigma}^+ \hat{c}_{l+1,\sigma}^{\phantom{+}} +
\hat{c}_{l+1,\sigma}^+ \hat{c}_{l,\sigma}^{\phantom{+}}
\right)
\left( 1-\hat{n}_{l+1,-\sigma}\right) 
\\[6pt]
\hat{T}_{\rm UHB} &=& (-t) \sum_{l,\sigma} (1+(-1)^l\delta) 
\hat{n}_{l,-\sigma}
\left(
\hat{c}_{l,\sigma}^+ \hat{c}_{l+1,\sigma}^{\phantom{+}} +
\hat{c}_{l+1,\sigma}^+ \hat{c}_{l,\sigma}^{\phantom{+}}
\right)
\hat{n}_{l+1,-\sigma} 
\\[6pt]
\hat{D}&=& \sum_l \hat{n}_{l,\uparrow}\hat{n}_{l,\downarrow}\\[6pt]
\hat{V} &=& \sum_l (\hat{n}_l-1)(\hat{n}_{l+1}-1) 
\end{eqnarray}
\end{mathletters}%
where 
$\hat{n}_{l,\sigma}=\hat{c}_{\l,\sigma}^+\hat{c}_{\l,\sigma}^{\phantom{+}}$
is the local density of $\sigma$-electrons and
$\hat{n}_{l}=\hat{n}_{l,\uparrow}+\hat{n}_{l,\downarrow}$
is the total local electron density.
$U\hat{D}$ is the Hubbard interaction between electrons on the same site,
$V\hat{V}$ is the nearest-neighbor interaction between charged objects
like double occupancies and holes, and
$\hat{T}_{\rm LHB}$ ($\hat{T}_{\rm UHB}$) describes the motion of
holes (double occupancies) in the lower (upper) Hubbard band.
The Harris-Lange model
is equivalent to the Hubbard model to order~$t(t/U)^{-1}$,
$t(t/U)^{0}$, and $t(V/U)^{0}$. 
Due to the Peierls distortion the electron transfer between two lattice sites
is modulated by~$\pm t\delta$.
We are only interested in the half-filled case where the
number of electrons~$N$ equals the even number of lattice sites~$L$.

\subsection{Band structure interpretation}

For~$V=0$ and in the absence of a lattice distortion the
exact eigenenergies of the Harris-Lange model
are the same as those of independent spinless Fermions moving
in two parallel bands separated by~$U$. 
For the case of linear optical absorption we may thus work with
the effective band structure Hamiltonian
\begin{equation}
\hat{H}_{\rm HL}^{\rm band} =  \sum_{|k|\leq \pi/a}\left[
(U+\epsilon(k)) \hat{n}_{k}^{u} + \epsilon(k) \hat{n}_{k}^{l} \right]
\label{effHL}
\end{equation}
with $\epsilon(k)=-2ta \cos(ka)$, and
$\hat{n}_{k}^{u}=\hat{u}_{k}^+  \hat{u}_{k}^{\phantom{+}}$,
$\hat{n}_{k}^{l}=\hat{l}_{k}^+  \hat{l}_{k}^{\phantom{+}}$
for our fermionic quasi-particles (chargeons)
in the upper and lower Hubbard band.
Their band width is~$W=4t$, the lattice spacing is~$a$, 
and the $k$-values of the first Brillouin zone
are spaced by~$\delta k=2\pi/(La)$.

The lattice distortion
results in a Peierls splitting of the upper and lower Hubbard band.
In this case the effective Hamiltonian for linear optical
absorption becomes  (Gebhard {\em et al.} II 1996)    
\begin{equation}
\hat{H}_{\rm HL}^{\rm dim,\, band} = 
 \sum_{|k|\leq \pi/(2a),\tau=\pm 1}\biggl[
(U+\tau E(k)) \hat{n}_{k,\tau}^{u} + \tau E(k) 
\hat{n}_{k,\tau}^{l} \biggr]  \; .
\label{bandhldim}
\end{equation}
with the dispersion relation
\begin{mathletters}
\begin{eqnarray}
E(k)&=&\sqrt{\epsilon(k)^2+\Delta(k)^2} \\[3pt]
\epsilon(k) &=& -2t \cos(ka) \\[3pt]
\Delta(k)&=& 2t\delta \sin (ka) 
\end{eqnarray}
\end{mathletters}%
where $\hat{n}_{k,\tau}^{u}$ ($\hat{n}_{k,\tau}^{l}$) 
are the number operators for
the quasi-particles for the upper ($\tau=+$) and lower ($\tau=-$)
Peierls subband in the upper~($u$) and lower~($l$) Hubbard band.

\subsection{Spin degrees of freedom at strong coupling: 
the Heisenberg model}
\label{spinhamilT}

At half-filling the ground state of the Harris-Lange model is~$2^L$-fold
degenerate. We are ultimately interested in the optical properties
of the (Peierls-distorted extended) Hubbard model at strong correlations 
for which the ground state for large~$U/W$ is unique. 
Consequently, we have to go to the next order
in the expansion of the Hubbard model
in~$W/U$ to lift the above degeneracy.
The Harris-Lange model is unsatisfactory in yet another aspect:
at half filling a finite lattice distortion
cannot be sustained within the model
because there is no gain in electronic energy.
The electronic part of the ground state energy is zero, irrespective
of the Peierls parameter~$\delta$.

A consistent treatment of the expansion in~$W/U$
will also produce corrections to the Harris-Lange model 
(Harris and Lange 1967); (van Dongen 1994)                   
which would render the problem intractable.
Thus we argue that the degeneracy of the ground state will
inevitable be lifted by a residual spin-spin interaction which
may have its origin in the itinerant exchange in the Hubbard model
(next term in the expansion in $W/U$) or in their direct exchange
which is not taken into account in the extended Hubbard model
where only the direct Coulomb terms were kept
(``Zero Differential Overlap Approximation''
(Kivelson, Su, Schrieffer, and Heeger 1987); (Wu, Sun, and Nasu 1987);   
(Baeriswyl, Horsch, and Maki 1988); (Gammel and Campbell 1988);
(Kivelson, Su, Schrieffer, and Heeger 1988);
(Campbell, Gammel, and Loh 1988); (Painelli and Girlando 1989);
(Campbell, Gammel, and Loh 1990)).
The additional ``Hubbard-$W$-terms'' might very well
be important in strongly-correlated narrow-band materials.

Thus we {\em assume\/} that the {\em charge\/} dynamics
is still governed by the Harris-Lange model.
In particular, we assume that the chargeons do not scatter from
the spinons.
For the {\em spin\/} dynamics we choose the (dimerized, anisotropic) 
Heisenberg Hamiltonian
with an anisotropy into the $z$-direction
\begin{equation}
\hat{H}_{\rm Heis} = \sum_{l} \left(1+(-1)^l\delta_S\right)
\left[ J_{\perp}\left(\hat{S}_l^x \hat{S}_{l+1}^x
+ \hat{S}_l^y \hat{S}_{l+1}^y \right) + J_{z}
\hat{S}_l^z \hat{S}_{l+1}^z
 - \frac{1}{4} \right]
\label{dimHeisHamilt}
\end{equation}
where~$\hat{\rm\bf S}_l$ is the vector operator for the spin at site~$l$,
$\hat{S}_l^+=\hat{S}_l^x+i\hat{S}_l^y=
\hat{c}_{l,\uparrow}^+\hat{c}_{l,\downarrow}^{\phantom{+}}$,
$\hat{S}_l^-=\hat{S}_l^x-i\hat{S}_l^y=
\hat{c}_{l,\downarrow}^+\hat{c}_{l,\uparrow}^{\phantom{+}}$,
and
$\hat{S}_l^z=(\hat{n}_{l,\uparrow}-\hat{n}_{l,\downarrow})/2$.
$J_{\perp}>0$ and $J_z>0$ are two generally different
antiferromagnetic coupling constants, 
and $0\leq \delta_S\leq 1$
determines the degree of spin
dimerization. It is well-known that a one-dimensional
spin-system that is coupled to the lattice degrees of freedom shows
the spin-Peierls effect, $\delta_S >0$.

If the itinerant exchange was responsible for the antiferromagnetic coupling
we would have~$J\equiv 
J_{\perp}=J_z=(1+\delta^2) 2t^2/U$ and $\delta_S=2\delta/(1+\delta^2)$.
In general, however, we do not assume a simple relation between~$J_{\perp}$,
$J_z$ and~$t$,
or~$\delta_S$ and~$\delta$.
We state, however, that $J_{\perp}$, $J_z$ 
are supposed to be {small\/} energy scales
compared to the Coulomb energies~$U$,~$V$. 
Its influence on the exact optical excitation energies will thus be neglected
in this work. This implies, for example, that  we cannot distinguish between
singlet and triplet excitons. 

\section{Optical absorption and effective current operator}
\label{optabssec}
\subsection{Optical conductivity and optical absorption}

The dielectric function~$\widetilde{\epsilon}(\omega)$
and the coefficient for the linear optical
absorption~$\widetilde{\alpha}(\omega)$ are
given by (Haug and Koch 1990)                 
\begin{mathletters}
\begin{eqnarray}
\widetilde{\epsilon}(\omega) &=& 1 +\frac{4\pi i \sigma(\omega)}{\omega}
\label{epssigma}\\[6pt]
\widetilde{\alpha}(\omega) &=&
\frac{4\pi {\rm Re}\{\sigma(\omega)\}}{n_b c}
\end{eqnarray}
\end{mathletters}%
where ${\rm Re}\{\ldots\}$ denotes the real part and
$n_b$ is the background refractive index. 
It is supposed to be frequency independent near a resonance.
Hence, the real part of the optical conductivity
directly gives the absorption spectrum of the system.

The standard result (Maldague 1977); (Mahan 1990)     
for the real part of the optical
conductivity in terms of the current-current correlation 
function~$\chi(\omega)$ is
\begin{eqnarray}
{\rm Re}\{ \sigma(\omega) \} &=&\frac{{\rm Im}\{\chi(\omega)\}}{\omega}
\\[6pt]
\chi(\omega) & =& \frac{{\cal N}_{\perp}}{La}
i \int_0^{\infty} dt e^{i\omega t} \langle
\left[\hat{\jmath}(t),\hat{\jmath}\right]_- \rangle
\end{eqnarray}
where~${\cal N}_{\perp}$ is the number of chains per unit area
perpendicular to the chain direction.

The current-current correlation function can be spectrally
decomposed in terms of exact eigenstates of the system as
\begin{equation}
\chi(\omega) = \frac{{\cal N}_{\perp}}{La}
\sum_n |\langle 0 | \hat{\jmath} | n\rangle|^2
\left[ \frac{1}{\omega +(E_n-E_0) +i\gamma} -
\frac{1}{\omega -(E_n-E_0) +i\gamma} \right] \; .
\label{decomp}
\end{equation}
Here, $|0\rangle$ is the exact ground state (energy $E_0$),
$|n\rangle$ are exact excited states (energy $E_n$),
and $\left|\langle 0 | \hat{\jmath} | n\rangle\right|^2$
are the oscillator strengths for optical transitions between them.
Although $\gamma =0^+$ is infinitesimal we may introduce $\gamma>0$
as a phenomenological broadening
of the resonances at $\omega = \pm(E_n-E_0)$.
The spectral decomposition of the real part of the optical conductivity
reads
\begin{equation}
{\rm Re}\{ \sigma(\omega) \} = \frac{{\cal N}_{\perp} \pi}{La \omega}
\sum_n \left| \langle 0 | \hat{\jmath} | n\rangle\right|^2
\left[ \delta\left( \omega
-(E_n-E_0)\right) -\delta\left( \omega +(E_n-E_0)\right) \right]
\label{speccomp}
\end{equation}
which is positive for all~$\omega$.

In the following we will always plot the dimensionless
reduced optical conductivity
\begin{equation}
\sigma_{\rm red}(\omega >0) =
\frac{\omega {\rm Re}\{\sigma(\omega>0)\}  }%
{{\cal N}_{\perp}a e^2 W  } \; .
\label{sigmared}
\end{equation}
Furthermore we replace the energy conservation~$\delta(x)$
by the smeared function
\begin{equation}
\widetilde{\delta}(x) = \frac{\gamma}{\pi(x^2+\gamma^2)} 
\end{equation}
to include effects of phonons and experimental resolution. 

\subsection{Effective current operator}

In  (Gebhard {\em et al.} II 1996)    
we showed that the current operator for an excitation from
the filled lower Hubbard band to the empty upper Hubbard band 
can be written as
\begin{eqnarray}
\hat{\jmath}_{{\rm inter},+}^{\rm H} &=&
-(itea) \sum_{l,\sigma}\left(1+(-1)^l \delta\right) \left(1+(-1)^l \eta\right)
\nonumber \\[3pt]
&& \phantom{-(itea)\sum}
\left[
\hat{n}_{l+1,-\sigma} \hat{c}_{l+1,\sigma}^+ \hat{c}_{l,\sigma}^{\phantom{+}} 
\left( 1- \hat{n}_{l,-\sigma} \right)
-
\hat{n}_{l,-\sigma} \hat{c}_{l,\sigma}^+ \hat{c}_{l+1,\sigma}^{\phantom{+}} 
\left( 1- \hat{n}_{l+1,-\sigma} \right)
\right] 
\end{eqnarray}
where $\eta=-|R_{l+1}-R_l-a|/a<0$ is the relative change of lattice distances
due to the Peierls distortion (Gebhard {\em et al.} I 1996).        
$\hat{\jmath}_{\rm inter,+}^{\rm H}$ creates
a neighboring pair of a double occupancy and a hole.

As a central result of (Gebhard {\em et al.} II 1996)
we found the band structure representation
of the current operator as
\begin{mathletters}
\begin{equation}
\hat{\jmath}_{\rm inter,+}^{\rm band}=
\sum_{|k|,|q|\leq \pi/a}
iea\epsilon(k) \hat{u}_{k+q/2}^+\hat{l}_{k-q/2}^{\phantom{+}}
\hat{x}_{q}^{\phantom{+}} \; .
\label{jhlbandpicture}
\end{equation}
for the translational invariant case, and
\begin{eqnarray}
\hat{\jmath}_{{\rm inter},+}^{\rm band,\, dim}&=&
\sum_{|q|,|k|\leq \pi/(2a)}\biggl\{ iea \epsilon(k) \Bigl[
\hat{u}_{k+q/2}^+ \hat{l}_{k-q/2}^{\phantom{+}}  
-  \hat{u}_{k+q/2+\pi/a}^+ \hat{l}_{k-q/2+\pi/a}^{\phantom{+}} 
\Bigr] \hat{x}_{q}^{\phantom{+}} \label{jinterband}\\[9pt]
&& \phantom{\sum_{|q|,|k|\leq \pi/(2a)}\biggl\{ }
+ea \frac{\Delta(k)}{\delta} \Bigl[
 \hat{u}_{k+q/2+\pi/a}^+ \hat{l}_{k-q/2}^{\phantom{+}}
- \hat{u}_{k+q/2}^+  \hat{l}_{k-q/2+\pi/a}^{\phantom{+}}
\Bigr] \hat{x}_{q+\pi/a}^{\phantom{+}}  \biggr\} \nonumber
\end{eqnarray}
\end{mathletters}%
for a Peierls-distorted lattice. The diagonalized form
of the interband current operator is given in 
(Gebhard {\em et al.} II 1996).                         

Since the excitation energy only depends on the charge configuration
the complicated spin problem could be hidden in the operators~$\hat{x}_q$
which are defined only in terms of their product,
{\arraycolsep=0pt\begin{eqnarray}
\hat{x}_q^{+}(\delta,\eta)\hat{x}_{q'}^{\phantom{+}}(\delta,\eta)
&=& 
\sum_{S_1^{\prime},\ldots S_{L-2}^{\prime}}
\frac{1}{L^2} \sum_{l,r} e^{i(ql-q'r)}
\bigl(1+\eta\delta +(-1)^l(\delta+\eta)\bigr)
\bigl(1+\eta\delta +(-1)^r(\delta+\eta)\bigr)
\nonumber \\[6pt]
&& \phantom{\sum_{S_1^{\prime},\ldots S_{L-2}^{\prime}}
\frac{1}{L^2} \sum_{l,r} e^{i(ql-q'r)} }
\langle 0 | S_1^{\prime},\ldots S_{l-1}^{\prime},
\left( \uparrow_{l}\downarrow_{l+1}-\downarrow_{l}\uparrow_{l+1}\right),
S_{l}^{\prime},\ldots S_{L-2}^{\prime} \label{xqxqprime} \rangle
\\[6pt]
&& \phantom{\sum_{S_1^{\prime},\ldots S_{L-2}^{\prime}}
\frac{1}{L^2} \sum_{l,r} e^{i(ql-q'r)} }
\langle S_{L-2}^{\prime},\ldots S_{r}^{\prime},
\left( \downarrow_{r+1} \uparrow_{r}-\uparrow_{r+1}\downarrow_{r}\right),
S_{r-1}^{\prime},\ldots S_{1}^{\prime} | 0 \rangle \; .
\nonumber
\end{eqnarray}
In practice, $q'=q$ or $q'=q+\pi/a$.}
In (Gebhard {\em et al.} II 1996)                         
we studied the ``hot-spin'' case where all $2^L$~spin configurations
were equally possible ground states. Now we will set up
the ``no-recoil'' approximation for the case of the (unique) ground state
of the Heisenberg model, equation~(\ref{dimHeisHamilt}).

\section{No-recoil approximation}
\label{norecoilsec}

\subsection{Basic ideas}

Although the eigenenergies and even the eigenstates of the Hubbard
model at strong coupling nicely display charge-spin 
separation (Ogata and Shiba 1990); (Parola and Sorella 1990)      
the two dynamical degrees of
freedom are coupled again by the current operator that
creates an optical excitation. 
The oscillator strength for such an excitation, however,
factorizes into a charge and a spin part. For the charge
part our band structure picture applies.
The spin part is determined by
a ground state correlation function for nearest-neighbor singlets
at all lattice distances, see below.
Hence, the calculation
of the optical absorption spectrum remains a difficult
many-particle problem even for a formally charge-spin separated
system. 

If we want to make further progress we have to make assumptions
on the behavior of the spin system. In~(Gebhard {\em et al.} II 1996)
we treated the
case of a degenerate spin background (``hot-spin case'')
where all spin configurations were equivalent.
Now we are interested in the more realistic case where the
optical excitation starts from the unique ground state of the
Heisenberg model, equation~(\ref{dimHeisHamilt}).

Our band structure interpretation suggests an analogy between our
strongly correlated Mott-Hubbard system and the situation
in an intrinsic semiconductor with a direct gap. 
The Hubbard bands for the charges
correspond to conduction and valence bands, while the spin degrees of
freedom play the role of phonons in a semiconductor or metal.
In both cases the energy scales are well separated, the spinon (phonon)
energy being much smaller than the chargeon (electron) energy.
The spin spectrum is gaped in the presence of dimerization ($\delta_S\neq 0$)
such that the spin excitations correspond to optical
rather than acoustical phonons in that case.
This appealing
analogy is also the basic idea behind the tJ-model approach to high-Tc
superconductivity (Zhang and Rice 1988); (Dagotto 1994).             

In electron-phonon systems
one may often {\em ignore\/ } phonon emission or absorption
during an excitation at low temperatures.
This is an approximation even at zero temperatures since
phonons can always be emitted even if phonon absorption is impossible.
Effectively there is a finite probability (``Debye-Waller factor'')
that there is no momentum transferred to the phonons
during an excitation by photons.
This constitutes the ``no-recoil approximation''.
It has many applications in solid state science (e.g.,
X-ray scattering, M\"{o}\ss bauer-effect),
and it is also very successfully applied
for the explanation of optical spectra of semiconductors 
(Haug and Koch 1990).                                       
Thus we are confident that a (modified) ``no-recoil approximation'' 
will also be
valid for the strongly correlated Hubbard model where chargeons and
spinons are energetically well separated.

\subsection{Formulation of the approximation}

To formulate the approximation for our case we will scrutinize
equation~(\ref{xqxqprime}) for the spin sector. Recall that the
charge sector has already been mapped onto a standard band structure picture. 
We apply
{\arraycolsep=0pt\begin{eqnarray}
| S_1^{\prime},\ldots S_{l-1}^{\prime},
\left( \uparrow_{l},\downarrow_{l+1}-\downarrow_{l},\uparrow_{l+1}\right),
S_{l}^{\prime},\ldots S_{L-2}^{\prime} \rangle
&=& \\[6pt]
&& \hspace*{-6pt} \hat{{\cal T}}_S^{(l-2)}\hat{{\cal T}}_{S'}^{-(l-2)}
| S_1^{\prime}
\left( \uparrow_{2},\downarrow_{3}-\downarrow_{2},\uparrow_{3}\right),
S_{2}^{\prime},\ldots S_{L-2}^{\prime} \rangle \nonumber
\end{eqnarray}
for both expectation values in equation~(\ref{xqxqprime})
where~$\hat{\cal T}_{S}$ shifts all spins by one site
and $\hat{\cal T}_{S'}$ shifts all spins
but those at sites~$2$ and~$3$.}

Furthermore, we use the identity
\begin{equation}
2 \sum_{S_2,S_3} \left(\frac{1}{4}-\hat{\rm\bf S}_2\hat{\rm\bf S}_{3} \right)
| S_2, S_3\rangle\langle S_3, S_2 |
\left(\frac{1}{4}-\hat{\rm\bf S}_2\hat{\rm\bf S}_{3} \right)
= | \left( \uparrow_2, \downarrow_3 - \uparrow_3, \downarrow_2\right)
\rangle
\langle \left( \downarrow_3, \uparrow_2 - \downarrow_2, \uparrow_3\right)|
\; .
\end{equation}
Now that we have eliminated the restrictions on the intermediate
spin sum we can carry it out and are left with a complicated
ground state expectation value,
\begin{mathletters}
\begin{eqnarray}
\hat{x}_q^{+}\hat{x}_{q'}^{\phantom{+}} &=& 2 \langle 0 |
\hat{Z}_{2,3}^{+}(q)
\left(\frac{1}{4}-\hat{\rm\bf S}_2\hat{\rm\bf S}_{3} \right)
\hat{Z}_{2,3}^{\phantom{+}}(q')
| 0 \rangle
\\[6pt]
\hat{Z}_{2,3}^{\phantom{+}}(q) &=&
\frac{1}{L} \sum_l e^{-iqla}
\hat{{\cal T}}_S^{(l-2)}\hat{{\cal T}}_{S'}^{-(l-2)}
\left( 1+ (-1)^l\delta\right)\left( 1+ (-1)^l\eta\right) \; .
\end{eqnarray}
\end{mathletters}%
Thus far the expressions are exact.
Note that $\hat{x}_q^+\hat{x}_{q'}^{\phantom{+}}$
is almost --but not quite-- the
double Fourier transform of the correlation function of nearest-neighbor
singlets at site~$l$ and~$r$.

To obtain further insight into the problem we first analyze
two special cases which can exactly be solved before we formulate the general
approximation.

\subsubsection{N\'{e}el state}

The N\'{e}el state is the ground state of the anisotropic Heisenberg model,
equation~(\ref{dimHeisHamilt}), for $J_{\perp}=0$ (Ising model),
\begin{equation}
| \hbox{AF}\rangle = \prod_{l=1}^{L/2}
\hat{c}_{2l,\uparrow}^+\hat{c}_{2l+1,\downarrow}^+|\hbox{vacuum}\rangle \; .
\end{equation}
It displays a perfect double-periodic structure, i.~e., if we shift all spins by
two lattice sites we recover $|\hbox{AF}\rangle$.
Hence, 
\begin{equation}
\hat{{\cal T}}_S^{(l-2)}\hat{{\cal T}}_{S'}^{-(l-2)} |\hbox{AF}\rangle
=\frac{1}{2} \left[\left(1+(-1)^l\right) + \left( 1 - (-1)^l\right) 
\hat{{\cal T}}_S^{\phantom{1}}
\hat{{\cal T}}_{S'}^{-1}\right] |\hbox{AF}\rangle \; .
\label{AFshift1}
\end{equation}
The wave function renormalization factors become
\begin{mathletters}
\label{AFshift2}
\begin{eqnarray}
\hat{Z}_{2,3}^{\phantom{+}}(q) |\hbox{AF}\rangle &=&
\frac{\delta_{q,0}+\delta_{q,\pi/a}}{2}|\hbox{AF}(q)\rangle
\\[6pt]
|\hbox{AF}(q)\rangle &=& \left( (1+\delta)(1+\eta) + e^{iqa}
(1-\delta)(1-\eta)
\hat{{\cal T}}_S^{\phantom{1}}\hat{{\cal T}}_{S'}^{-1}
\right) |\hbox{AF}\rangle
\; .
\end{eqnarray}
\end{mathletters}%
Thus we may write
\begin{equation}
\left. \hat{x}_q^+\hat{x}_{q'}^{\phantom{+}} \right|_{\rm AF}
= \frac{1}{2} \left(\delta_{q,0}+\delta_{q,\pi/a} \right)
\left(\delta_{q',0}+\delta_{q',\pi/a} \right)
\langle \hbox{AF}(q) |
\frac{1}{4}-\hat{\rm\bf S}_2\hat{\rm\bf S}_{3} |
\hbox{AF}(q') \rangle \; .
\label{neuneuneuAF}
\label{AFshift3}
\end{equation}
For later use we define
\begin{mathletters}
\label{theZs}
\begin{eqnarray}
Z_0^{\rm AF} &=& \langle \hbox{AF}(0) |
\frac{1}{4}-\hat{\rm\bf S}_2\hat{\rm\bf S}_{3} |
\hbox{AF}(0) \rangle \\[6pt]
Z_{\pi}^{\rm AF} &=& \langle \hbox{AF}(\pi/a) |
\frac{1}{4}-\hat{\rm\bf S}_2\hat{\rm\bf S}_{3} |
\hbox{AF}(\pi/a) \rangle \\[6pt]
Z_M^{\rm AF} &=& \left[ \langle \hbox{AF}(0) |
\frac{1}{4}-\hat{\rm\bf S}_2\hat{\rm\bf S}_{3} |
\hbox{AF}(\pi/a) \rangle + {\rm h.c.}\right]
\end{eqnarray}
\end{mathletters}%
such that $\left. \hat{x}_0^+\hat{x}_{0}^{\phantom{+}} \right|_{\rm AF}
=Z_0^{\hbox{\scriptsize AF}}/2$,
$\left. \hat{x}_{\pi/a}^+\hat{x}_{\pi/a}^{\phantom{+}} \right|_{\rm AF}
=Z_{\pi}^{\hbox{\scriptsize AF}}/2$,
and $\left. \left(\hat{x}_{\pi/a}^+\hat{x}_{0}^{\phantom{+}} +
\hat{x}_{0}^+\hat{x}_{\pi/a}^{\phantom{+}} \right) \right|_{\rm AF}
=Z_M^{\hbox{\scriptsize AF}}/2$.
Explicitly,
\begin{equation}
\begin{array}{@{}rcl@{}}
Z_0^{\hbox{\scriptsize AF}} &=& 2(\delta+\eta)^2\\
Z_{\pi}^{\hbox{\scriptsize AF}} &=& 2(1+\delta\eta)^2 \\
Z_M^{\hbox{\scriptsize AF}} &=& 4(\delta+\eta)(1+\delta\eta)
\end{array}
\; .
\end{equation}
The N\'{e}el state dominantly
provides the momentum $\Delta q_S=\pi/a$
to the charge system during an optical excitation since 
$Z_{\pi}^{\hbox{\scriptsize AF}} >
Z_M^{\hbox{\scriptsize AF}}>
Z_0^{\hbox{\scriptsize AF}}$ 
for generic values for $\delta$, $\eta$.

We note that for $\delta=\eta=0$ only $Z_{\pi}^{\hbox{\scriptsize AF}}$ contributes.
This implies that for a N\'{e}el state one of the two Hubbard bands can be thought as 
being shifted by $\pi/a$ which results in vertical transitions
between {\em antiparallel\/} 
bands as for non-interacting electrons in a 
Peierls-distorted lattice (Gebhard {\em et al.} I 1996).

\subsubsection{Dimer state}

The ground state of the fully dimerized Heisenberg model, 
equation~(\ref{dimHeisHamilt})
for $J_{\perp}=J_z\equiv J$ and $\delta_S=1$, is the dimer state
\begin{equation}
| \hbox{DIM}\rangle = \prod_{l=1}^{L/2} \sqrt{\frac{1}{2}}
\left( \hat{c}_{2l,\uparrow}^+\hat{c}_{2l+1,\downarrow}^+
- \hat{c}_{2l,\downarrow}^+\hat{c}_{2l+1,\uparrow}^+\right)|\hbox{vacuum}
\rangle \; .
\end{equation}
Again, the dimer state is invariant under a shift of all spins by two lattice sites.
Hence, equations~(\ref{AFshift1}), (\ref{AFshift2}),
(\ref{AFshift3}), and (\ref{theZs}) analogously
hold for the dimer state. 
In particular,
\begin{mathletters}
\begin{eqnarray}
\left(\frac{1}{4}-\hat{\rm\bf S}_2\hat{\rm\bf S}_{3}\right) |\hbox{DIM}\rangle&= &
|\hbox{DIM}\rangle \\[3pt]
\left(\frac{1}{4}-\hat{\rm\bf S}_2\hat{\rm\bf S}_{3} \right)
\hat{{\cal T}}_S^{\phantom{1}}\hat{{\cal T}}_{S'}^{-1}
|\hbox{DIM}\rangle&= & -\frac{1}{2} 
|\hbox{DIM}\rangle
\end{eqnarray}\end{mathletters}%
holds such that the $Z$-factors for the dimer state become
\begin{equation}
\begin{array}{@{}rcl@{}}
Z_0^{\hbox{\scriptsize DIM}}
&=& \left[1+\delta\eta +3(\delta+\eta)\right]^2/4\\
Z_{\pi}^{\hbox{\scriptsize DIM}}
&=&\left[3(1+\delta\eta) +\delta+\eta\right]^2/4\\
Z_M^{\hbox{\scriptsize DIM}} &=& \left[1+\delta\eta +3(\delta+\eta)\right]
\left[3(1+\delta\eta) +\delta+\eta\right]/2
\end{array}
\; .
\end{equation}
We see again that momentum transfer by $\Delta q_S=\pi/a$ dominates in
the dimer state but that there is zero momentum transfer 
even for $\delta=\eta=0$, in contrast to the N\'{e}el state.

\subsubsection{General case}

The above examples are limiting cases for the ground state
of the general Heisenberg model in equation~(\ref{dimHeisHamilt}).
The correlation function for nearest-neighbor spin singlets 
is long-ranged ordered such that the spin system
only allows for momentum transfers~$\Delta q_S=0, \pi/a$.
It has recently been shown by (Talstra, Strong, and Anderson 1995)
that this order persists in the standard Heisenberg model
($J_{\perp}=J_z=J$, $\delta_S=0$)  as ``hidden'' long-range order
although their is neither dimer nor antiferromagnetic
order in the ground state.

These observations make us confident that the following
``no-recoil'' approximation will give the
coherent features of the optical absorption
spectrum for all values of $J_{\perp}/J_z>0$
and $\delta_S$ in the dimerized Heisenberg model.
\begin{equation}
\hat{{\cal T}}_S^{l}\hat{{\cal T}}_{S'}^{-l} |0\rangle
=
\left[ w_{\rm DW} \left(\frac{1+(-1)^l}{2}\right)
+
\overline{w_{\rm DW}} \left(\frac{1-(-1)^l}{2}\right)
\hat{{\cal T}}_S^{\phantom{1}}\hat{{\cal T}}_{S'}^{-1}
\right]|0\rangle
\quad + {\rm  rest} \; .
\label{norecoildim}
\end{equation}
Due to the ``hidden'' long-range order of spin singlets
in the ground state we expect that we need two different
``Debye-Waller factors'' $w_{\rm DW}$, $\overline{w_{\rm DW}}$
as in the cases of the N\'{e}el and dimer state where they both were
unity. In general, the square of their absolute value is smaller than one
as can be checked from sum rules.
It is clear that more elegant approximations than this can be designed,
e.g., an~$l$-dependent factor~$w_{\rm DW}(l)$ could be introduced
to mimic a finite correlation length for
finite temperatures as in (Gebhard {\em et al.} II 1996).
We will not follow the latter ideas here.

The wave function renormalization factors become
\begin{mathletters}
\begin{eqnarray}
\hat{Z}_{2,3}^{\phantom{+}}(q) |0\rangle &=&
\frac{\delta_{q,0}+\delta_{q,\pi/a}}{2}|\Psi(q)\rangle
\quad + {\rm  rest} \\[6pt]
|\Psi(q)\rangle &=& \left( w_{\rm DW}(1+\delta)(1+\eta) + e^{iqa}
\overline{w_{\rm DW}}(1-\delta)(1-\eta)
\hat{{\cal T}}_S^{\phantom{1}}\hat{{\cal T}}_{S'}^{-1}
\right) |0\rangle
\; .
\end{eqnarray}
\end{mathletters}%
Thus we may write
\begin{equation}
\left. \hat{x}_q^+\hat{x}_{q'}^{\phantom{+}} \right|_{\rm coh}
= \frac{1}{2} \left(\delta_{q,0}+\delta_{q,\pi/a} \right)
\left(\delta_{q',0}+\delta_{q',\pi/a} \right)
\langle \Psi(q) |
\frac{1}{4}-\hat{\rm\bf S}_2\hat{\rm\bf S}_{3} |
\Psi(q') \rangle \; .
\label{neuneuneu}
\end{equation}
In the presence of a Peierls distortion there is no recoil from the spin system 
since the reciprocal lattice vector is given by~$Q=\pi/a$.
Even if the lattice distortion was absent, however, the 
translational symmetry is broken by the ``hidden'' long-range order
in the nearest-neighbor singlet correlation function.
The coherent part of the spin contribution can then be expressed
with the help of
$\left. \hat{x}_0^+\hat{x}_{0}^{\phantom{+}} \right|_{\rm coh}
=Z_0/2$,
$\left. \hat{x}_{\pi/a}^+\hat{x}_{\pi/a}^{\phantom{+}} \right|_{\rm coh}
=Z_{\pi}/2$,
and $\left. \left(\hat{x}_{\pi/a}^+\hat{x}_{0}^{\phantom{+}} +
\hat{x}_{0}^+\hat{x}_{\pi/a}^{\phantom{+}} \right) \right|_{\rm coh}
=Z_M/2$.

\section{Optical conductivity in the Hubbard model at strong coupling}
\label{sechubb}
\label{optabsHubbard}

In~(Gebhard {\em et al.} II 1996)
we calculated the optical absorption for the case of an
incoherent spin background. In the present case we only have to
replace the expressions for 
the operator products~$\hat{x}_q^+\hat{x}_{q'}^{\phantom{+}}$ by
those for the ``no-recoil approximation'',
equations~(\ref{norecoildim}) and~(\ref{neuneuneu}).

\subsection{Band case: $V=0$}

Since the ground state implicitly breaks the translational symmetry
of our system even for $\delta=\eta=0$ we directly approach
the case of non-vanishing $\delta$, $\eta$.

For the dimerized case we have to diagonalize the current operator,
equation~(\ref{jinterband}), in terms of the quasi-particle operators
for the respective Peierls bands.
The result for the optical conductivity can be written as
(Gebhard {\em et al.} II 1996)                             
\begin{mathletters}
\label{monstersigma}
\begin{equation}
{\rm Re}\{\sigma(\omega >0, \delta,\eta) \}
= \frac{\pi {\cal N}_{\perp}}{L a \omega}
\sum_{\tau,\tau'=\pm 1} \sum_{|q|,|k| \leq \pi/(2a)}
\left| \lambda_{\tau,\tau'} (k,q)\right|^2
\delta(\omega - E_{\tau,\tau'}(k,q))
\end{equation}
with the absorption energies between the respective Peierls subbands
\begin{equation}
E_{\tau,\tau'}(k,q) = U +\tau' E(k+q/2)-\tau E(k-q/2) \; .
\label{Etautauprime}
\end{equation}
\end{mathletters}%
The transition matrix elements are given by 
$\lambda_{+,+}(k,q)=-\lambda_{-,-}(k,q)$, $\lambda_{-,+}(k,q)=
\lambda_{+,-}(k,q)$, and
\begin{mathletters}
\label{thelambdas}
\begin{eqnarray}
\lambda_{+,+}(k,q) &=& iea \left[ \epsilon(k)
(\alpha_{+}^{\phantom{*}}\alpha_{-}^*
-\beta_{+}^{\phantom{*}}\beta_{-}^*) \hat{x}_q^+
+\frac{\Delta(k)}{\delta}
(\alpha_{+}^{\phantom{*}}\beta_{-}^*
+\beta_{+}^{\phantom{*}}\alpha_{-}^*) 
\hat{x}_{q+\pi/a}^+ \right] \\[6pt]
\lambda_{+,-}(k,q) &=& iea \left[ - \epsilon(k)
(\alpha_{+}^{\phantom{*}}\beta_{-}^*
+\beta_{+}^{\phantom{*}}\alpha_{-}^*) \hat{x}_q^+
+\frac{\Delta(k)}{\delta}
(\alpha_{+}^{\phantom{*}}\alpha_{-}^*
-\beta_{+}^{\phantom{*}}\beta_{-}^*)
\hat{x}_{q+\pi/a}^+  \right] \; .
\end{eqnarray}
\end{mathletters}%
Here we used the short-hand notation
$\alpha_{\pm}=\alpha_{k\pm q/2}$ etc.\ for the mixing amplitudes
in the standard Bogoliubov transformation
(Gebhard {\em et al.} I and II 1996).   

For the coherent part of the optical absorption we only need their
values at~$q=0$. In this case,
$\alpha_k^2-\beta_k^2=-\epsilon(k)/E(k)$ and
$2 \alpha_k\beta_k =-\Delta(k)/E(k)$.
We thus find
$\left. |\lambda_{+,\pm }(k,q)|^2\right|_{\rm coh}
= \delta_{q,0} |\lambda_{+,\pm}(k)|^2$
with
\begin{mathletters}
\begin{eqnarray}
|\lambda_{+,+}(k)|^2 &=& \frac{(ea)^2}{2}
\left[
Z_0 \frac{\epsilon(k)^4}{E(k)^2} +
\delta^2 Z_{\pi} \frac{\Delta(k)^4}{\delta^4 E(k)^2}
+ \delta Z_M \left(\frac{\epsilon(k)\Delta(k)}{\delta E(k)}\right)^2\right]
\\[6pt]
|\lambda_{+,-}(k)|^2 &=& \frac{(ea)^2}{2}
\left( \frac{\epsilon(k)\Delta(k)}{\delta E(k)} \right)^2
\left[ \delta^2 Z_0+Z_{\pi}-\delta Z_M\right]
\; .
\end{eqnarray}
\end{mathletters}%
Each of the two quantities above can be expressed in terms of an
expectation value of the projection operator
$(1/4-\hat{\rm\bf S}_2\hat{\rm\bf S}_{3})$. Hence
we can be sure that the result is positive or zero.

For vanishing nearest-neighbor interaction we may directly use
equation~(\ref{monstersigma}) to arrive at
\begin{eqnarray}
{\rm Re}\{\sigma_{\rm coh}(\omega >0, \delta,\eta) \}
& = & \frac{\pi {\cal N}_{\perp}}{a \omega}
\biggl[  \frac{2}{L} \sum_{|k| \leq \pi/(2a)} \delta(\omega-U)
\left| \lambda_{+,+} (k)\right|^2
\\[6pt]
&& 
+ \frac{1}{L} \sum_{|k| \leq \pi/(2a)}
\left| \lambda_{+,-} (k)\right|^2 
\Bigl( \delta(\omega - U- 2E(k)) +\delta(\omega - U+ 2E(k)) \Bigr)\biggr]
\; . \nonumber
\end{eqnarray}
For later use we define the abbreviations
\begin{mathletters}
\label{handx}
\begin{eqnarray}
H_1&=& \frac{16}{W^2 L}\sum_{|k|\leq \pi/(2a)} \frac{\epsilon(k)^4}{E(k)^2}
= 1-\delta^2 H_3\\[6pt]
H_2&=& \frac{16}{W^2 L}\sum_{|k|\leq \pi/(2a)}
\frac{\Delta(k)^4}{\delta^4 E(k)^2}
= \frac{1}{\delta^2}\left( 1 -H_3\right)\\[6pt]
H_3&=& \frac{16}{W^2 L}\sum_{|k|\leq \pi/(2a)}
\left( \frac{\epsilon(k)\Delta(k)}{\delta E(k)}\right)^2
= \left(\frac{1}{1+|\delta|}\right)^2
\\[6pt]
X(\omega) &=& -\frac{1}{\pi}\frac{16}{W^2 L}\sum_{|k|\leq \pi/(2a)}
\left( \frac{\epsilon(k)\Delta(k)}{\delta E(k)}\right)^2
\frac{\omega-U}{(\omega-U)^2-4E(k)^2} \; .
\end{eqnarray}
\end{mathletters}%
For vanishing nearest-neighbor interaction we only need the
imaginary part of the function~$X(\omega)$ which can easily be calculated.
The real part of the coherent optical conductivity finally becomes
\begin{eqnarray}
{\rm Re}\{\sigma_{\rm coh}(\omega >0, \delta,\eta) \}
&=& \frac{\pi {\cal N}_{\perp}(Wea)^2}{ 16 a\omega}
\biggl[ \left(Z_0 H_1+\delta^2 Z_{\pi}H_2+\delta
Z_MH_3\right) \delta(\omega-U)
\label{whynotnumbered}
\\[6pt]
&& 
+ \frac{2(\delta^2 Z_0+Z_{\pi}-\delta Z_M)
\sqrt{\left[(\omega-U)^2-(W\delta)^2\right]\left[W^2-(\omega-U)^2\right]}
}{
\pi W^2(1-\delta^2)^2|\omega-U|} \nonumber
\end{eqnarray}
in the regions~$\omega=U$ and $W\delta\leq |\omega-U|\leq W$.
Thus the generic coherent features are (i)~a $\delta$-peak at $\omega=U$
which results from vertical transitions between
the parallel Hubbard bands ($\Delta q_C=\Delta q_S=0$), and 
(ii)~a band-to-band transition feature for $W\delta\leq |\omega-U|\leq W$
which results from vertical transitions between
two antiparallel bands ($\Delta q_C=\Delta q_S=\pi/a$).

To check the sum rule we integrate over the reduced coherent
optical conductivity and obtain
\begin{eqnarray}
\int_0^{\infty} \frac{d\omega}{W} \sigma_{\rm red, coh}(\omega,\delta,\eta)
&=& \frac{\pi}{16}\left(Z_0+Z_{\pi}\right) \\[3pt]
&=& \frac{\pi}{4}\left[
|w_{\rm DW}|^2 (1+\delta)^2 (1+\eta)^2 C_S^{\rm even}
+ |\overline{w}_{\rm DW}|^2 (1-\delta)^2 (1-\eta)^2 C_S^{\rm odd}
\right]\nonumber
\end{eqnarray}
where $C_S^{\rm even,odd}$ is 
nearest-neighbor spin-correlation function on even/odd sites, see
equation~(\ref{CSevenCSodd}) of the appendix.
The comparison with the sum rule, equation~(\ref{sumrulesigmared}),
shows that~$|w_{\rm DW}|^2\leq 1$ and
$|\overline{w_{\rm DW}}|^2 \leq 1$ have to hold
which indeed allows to interpret them as
Debye-Waller factors.

For $\delta=1$, $\eta=0$ ($L/2$ independent two-site systems)
we have $Z_0=Z_{\pi}=Z_M/2=8 |w_{\rm DW}|^2
C_S^{\rm even}$. 
The contribution of the band vanishes again,
and  we obtain a single peak at $\omega=U$ with the coherent
fraction $|w_{\rm DW}|^2=1$ of the total oscillator strength.

The case of a vanishing Peierls distortion ($\delta=\eta=0$) is
particularly interesting. Recall that in the case
of an incoherent spin background (``hot-spin'' case) 
(Gebhard {\em et al.} II)
we observed a logarithmic divergence at $\omega=U$.
Now we find from
equation~(\ref{whynotnumbered}) that
\begin{equation}
{\rm Re}\{\sigma_{\rm red, coh}(\omega >0) \}
= \frac{1}{16}\left[W\pi Z_0 \delta(\omega-U)
+ 2Z_{\pi}\sqrt{1-\left( \frac{\omega-U}{W}\right)^2}\, \right]
\end{equation}
in which the proper $Z$-values for $\delta=\eta=0$ have to be inserted.
In general, the absorption consists of a $\delta$-peak at $\omega=U$
and a semielliptic contribution which come from vertical transitions
between the parallel ($\Delta q_C=\Delta q_S=0$) and
antiparallel ($\Delta q_S=\Delta q_C=\pi/a$) bands.

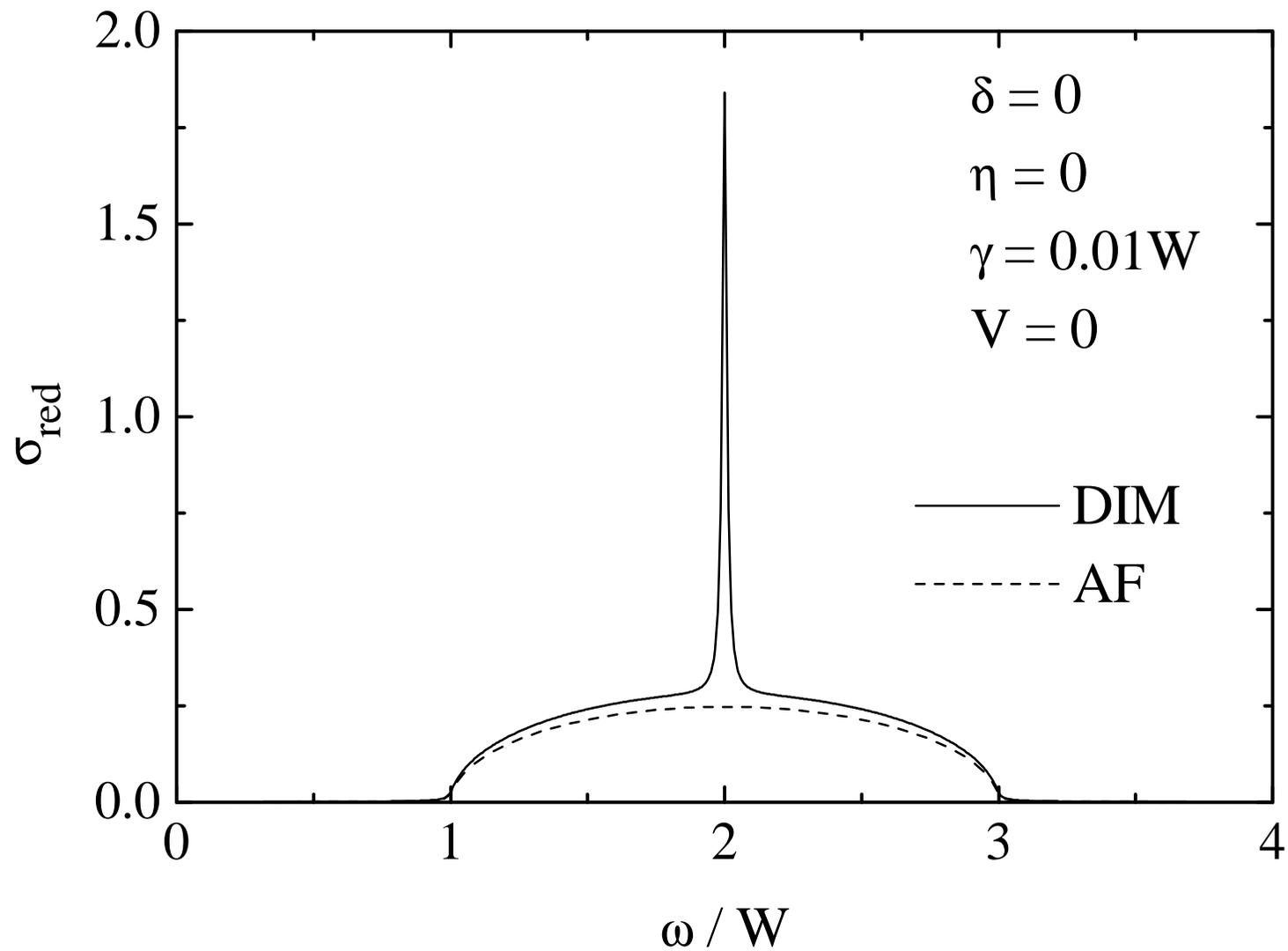
\begin{figure}[th]
\caption{Reduced optical conductivity,
$\sigma_{\rm red}(\omega >0)$,
for the dimer and the N\'{e}el state for $U=2W$. 
A broadening of~$\gamma=0.01W$ has been included.}
\label{heisfig00}
\end{figure}

For the N\'{e}el state $Z_0^{\hbox{\scriptsize AF}}(\delta=\eta=0)=0$, and
$Z_{\pi}^{\hbox{\scriptsize AF}}(\delta=\eta=0)=2$, and the result
of (Lyo and Galinar 1977); (Lyo 1978)                 
is recovered. It is {\em only\/} the N\'{e}el state
that suppresses the physics of the parallel Hubbard bands for $\delta=\eta=0$.
If we had only considered this state as reference state, the physics
of the parallel Hubbard bands would have been missed.
The dimer state with 
$Z_0^{\hbox{\scriptsize DIM}}(\delta=\eta=0)=1/4$, and
$Z_{\pi}^{\hbox{\scriptsize DIM}}(\delta=\eta=0)=9/4$
does show the generic features. The exact results for both states
are depicted in figure~\ref{heisfig00}.

For nonzero lattice dimerization ($\delta\neq 0$, $\eta\neq 0$)
even the N\'{e}el state reproduces the generic situation.
The coherent peak at $\omega=U$ is not at all a consequence
of the lattice distortion but a consequence of the
parallel Hubbard bands. This has been overlooked in
previous analytical investigations 
(Lyo and Galinar 1977); (Lyo 1978); (Galinar 1979).                   
Numerical calculations (Campbell, Gammel, and Loh 1989)    
for $\delta=\eta=0$ 
essentially give the results from the N\'{e}el ground state
since this state dominates for small system sizes.
Small traces of a peak at $\omega=U$ may have been
washed out by the adopted smoothing procedure.
For non-zero lattice distortion, however, the
$\delta$-resonance at $\omega=U$ becomes clearly visible
even in the numerical simulations.

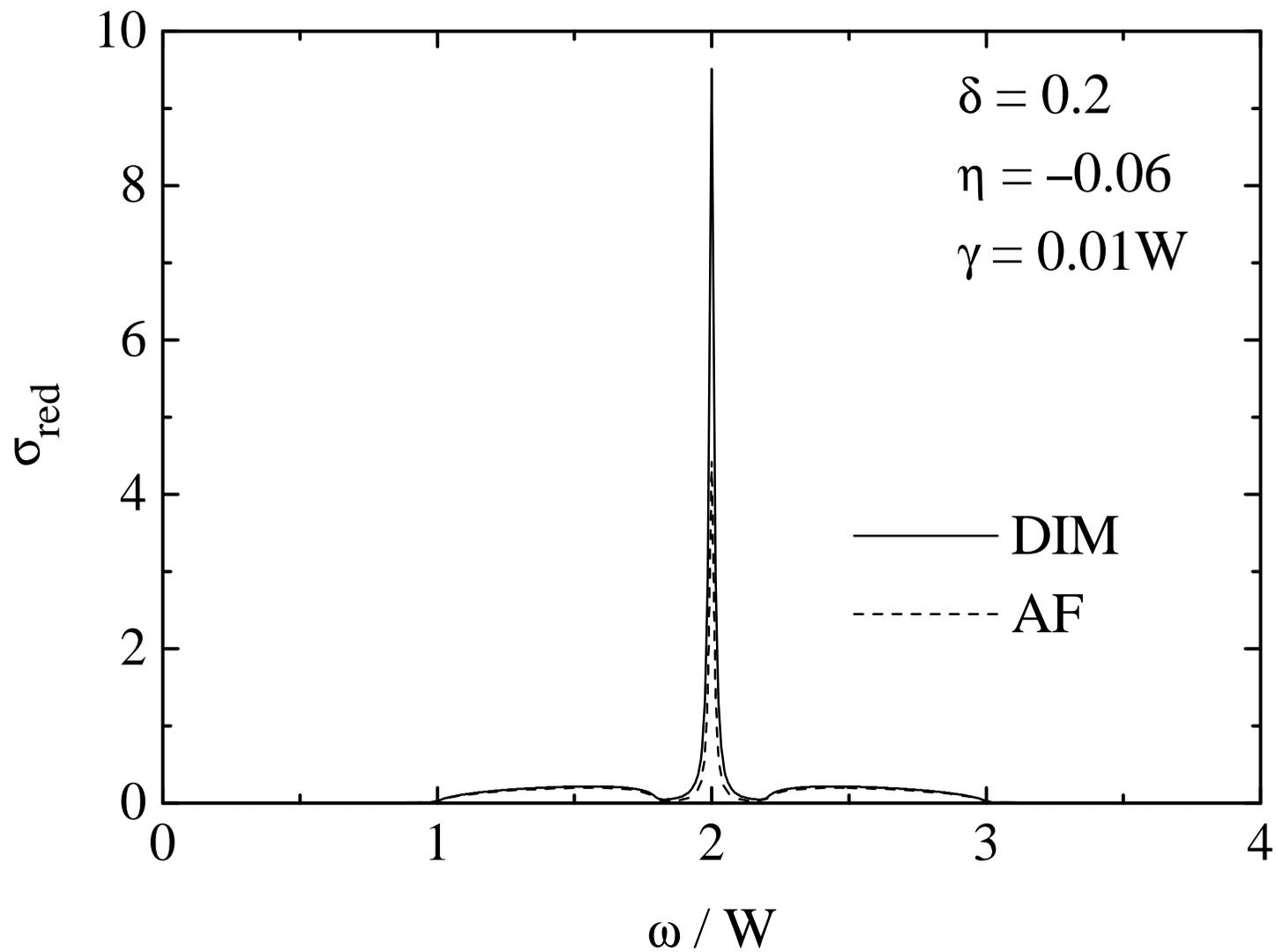
\begin{figure}[th]
\caption{Reduced optical conductivity,
$\sigma_{\rm red}(\omega >0, \delta,\eta)$,
for the dimer and the N\'{e}el state
for $U=2W$ in the presence of a lattice distortion, $\delta=0.2$, 
$\eta=-0.06$. 
A broadening of~$\gamma=0.01W$ has been included.}
\label{heisfig10}
\end{figure}

The exact results for the optical absorption
of the N\'{e}el and dimer state in the presence of
a lattice distortion are shown in figure~\ref{heisfig10}.
For realistic states the $Z$-factors will not be too different
such that the figure should reproduce the generic situation
for the Peierls-distorted Hubbard model 
at strong correlations and low temperatures.
This observation further supports our ``no-recoil approximation''.
Figure~\ref{heisfig10} has to be compared to figure~4 of
(Gebhard {\em et al.} II 1996).
It is seen that the linear absorption
at the threshold~$\omega=U-W$ now shows a square-root behavior.
This is also the case in the ``hot-spin'' case
for large enough~$\delta$ when the Peierls gap
has not been smeared out.
It is again seen that the significant features of the absorption spectrum
have not changed much when we go from the Harris-Lange model to the
strongly correlated Hubbard model.

\subsection{Exciton case: $V\neq 0$}

In the presence of a nearest-neighbor interaction we have to
solve an integral equation when the lattice is distorted.
Note that only the absolute position of the resonances will
be determined by the Coulomb parameter~$U/t$ while their relative
position depends on $V/t$. For illustrative purposes we tune $V/t$ 
independently of~$U/t$ and put aside the question of
the stability of the ground state
against the formation of a charge density wave.

The calculations are lengthy and will not be redone here,
see appendix~B of (Gebhard {\em et al.} II 1996)
for details. The result can be expressed in terms
of two operator-valued functions~$G_{1,2}(q)$,
\begin{mathletters}
\label{CapitalGred}
\begin{eqnarray}
G_1(q)&=&
itea \left[ \hat{x}_q^+ F_1(q) - \hat{x}_{q+\pi/a}^+ F_3(q)\right]\\[6pt]
G_2(q)&=&
itea \left[ -\hat{x}_q^+ F_3(q) + \hat{x}_{q+\pi/a}^+ F_2(q)\right]\; ,
\end{eqnarray}\end{mathletters}%
and three $q$-dependent functions $F_{1,2,3}(q)$ as
{\arraycolsep=0pt\begin{eqnarray}
{\rm Re}\{\sigma(\omega >0, V, \delta,\eta) \}
&=&
{\rm Re}\{\sigma(\omega >0,\delta,\eta) \}
+ \frac{2V{\cal N}_{\perp}}{a\omega}  \label{mostimportant}
\\[6pt]
&&{\rm Im}\Biggl\{ \sum_{|q| \leq \pi/(2a) }
\frac{1}{(1+VF_1)(1+VF_2)-(VF_3)^2}
\biggl[G_1^{\phantom{+}}G_1^+ + G_2^{\phantom{+}}G_2^+
\nonumber
\\[6pt]
&& \phantom{ {\rm Im}\biggl\{ \sum_{|q| \leq \pi/(2a) }   }
+V \left(
G_1^{\phantom{+}}G_1^+F_2 + G_2^{\phantom{+}}G_2^+F_1
+ (G_1^{\phantom{+}}G_2^+ + G_2^{\phantom{+}}G_1^+)F_3
\right)\biggr]\Biggr\} \; .
\nonumber\end{eqnarray}
In the ``no-recoil approximation'' we only need the $q=0$
value of the functions $F_{1,2,3}$.
We have}
\begin{mathletters}
\begin{eqnarray}
F_1(q=0)&=& \frac{H_1}{\omega-U} - \pi\delta^2 X(\omega) \\[3pt]
F_2(q=0)&=& \frac{\delta^2 H_2}{\omega-U} - \pi X(\omega) \\[3pt]
F_3(q=0)&=& -\delta \left[
\frac{H_3}{\omega-U} + \pi X(\omega)\right] \; ,
\end{eqnarray}
\end{mathletters}%
see eqs.~(\ref{handx}).
The definitions~(\ref{theZs}) allow us
to reduce equation~(\ref{mostimportant}) to
\begin{eqnarray}
{\rm Re}\{\sigma_{\rm coh}(\omega >0, V, \delta,\eta) \}
&=&
{\rm Re}\{\sigma_{\rm coh}(\omega >0,\delta,\eta) \}
\nonumber \\[6pt]
&&
+ \frac{V{\cal N}_{\perp}(Wea)^2}{16 a\omega}
{\rm Im}\Biggl\{
\frac{1}{(1+VF_1)(1+VF_2)-(VF_3)^2}
\label{thefinalresultHubb}
\\[6pt]
&& \phantom{+}
\biggl[ Z_0(F_1^2+F_3^2)+Z_{\pi}(F_2^2+F_3^2)-Z_M(F_1+F_2)F_3
\nonumber \\[3pt]
&& \phantom{+\biggl[ }
+V\left(F_1F_2-F_3^2\right)\left(Z_0F_1+Z_{\pi}F_2-Z_MF_3\right)\biggr]
 \Biggr\}\; .
 \nonumber
\end{eqnarray}
Note that the {\em positions\/} of the excitons are determined
by the zeros of the denominator which is a function of $\omega$, $V$,
and $\delta$ but is independent of the $Z$-factors and $\eta$.
The oscillator strength of an exciton, however, strongly
depends on the $Z$-factors.

For the translational invariant case, 
$\delta=\eta=0$, equation~(\ref{thefinalresultHubb})
can be simplified to ($F_3=0$, $F_1=1/(\omega-U)$, $F_2=-\pi X(\omega)$)
\begin{equation}
{\rm Re}\{\sigma(\omega >0,V) \}
= - \frac{(Wea)^2{\cal N}_{\perp}}{16 a\omega} 
{\rm Im}\left\{ \frac{Z_0F_1}{1+VF_1} + 
\frac{Z_{\pi}F_2}{1+VF_2} \right\} \; .
\end{equation}
{}From the first part we see that the peak at $\omega=U$ for $V=0$
becomes a $q=0$-exciton at frequency $\omega_0=U-V$.
For the N\'{e}el state this exciton is absent
while it is present {\em in all generic cases}, e.~g., in
the dimer state. The second part describes the formation of the 
$q=\pi/a$-exciton from the semielliptic ``band absorption'' part for $V=0$.
The formation of this exciton leads to a redshift in the
``band absorption'' part as already noticed by
(Galinar 1979).                   

The exact result for $\delta=\eta=0$ for the dimer state 
is displayed in figure~\ref{heisfig01} for various values of $V/t$.
For $V>W/2$ the $q=\pi/a$-exciton becomes a true bound state
at $\omega_{\pi/a}=U-V-W^2/(4V)$ below
the band absorption edge at $\omega_{\rm edge}=U-W$.
For large enough~$V/t$ the fact that~$Z_{\pi} > Z_0$ implies
that the $q=\pi/a$-exciton resonance is stronger than the one from 
the $q=0$-exciton. Since they are only separated 
by an energy difference $\delta\omega=W^2/(4V)$
thermal and disorder broadening could merge
the two exciton lines into a single asymmetric absorption
line. 

The general case of the optical absorption
of the dimer state in the presence of a
lattice distortion is shown in figure~\ref{heisfig11}.
The N\'{e}el state results in
a very similar curve  such that we are confident
that it reproduces the generic features for 
the coherent absorption in the Hubbard model at strong correlations.
At $V=0$ we had the resonance at $\omega=U$ and  
a Peierls-split ``band absorption''
in the range $W\delta \leq |\omega-U|\leq W$.
For moderate~$V/t$ the two bands evolve into two peaks such that
three lines are visible. The peak at energy $\omega_{\pi/a}=U-V-W^2/(4V)$
corresponds to the $q=\pi/a$-exciton while the dominant peak is 
the~$q=0$-exciton at~$\omega_0=U-V$. The peak near $\omega=U$ corresponds
to the {\em anti-bound\/} exciton which quickly looses oscillator
strength as $V/t$ increases.
For large~$V/t$ the $q=0$- and $q=\pi/a$-excitons 
dominate the absorption spectrum
with a preference of the $q=\pi/a$-exciton since $Z_{\pi}>Z_0$.
In addition, a weak structure remains near $\omega=U$
which does not play a role for the optical absorption but
may become visible in electroabsorption,
i.e., optical absorption in the presence of a static electrical field.

The dimer state underestimates the factors~$Z_{\pi}$, $Z_M$ as compared
to $Z_0$. For the N\'{e}el state the strength of the
$q=0$-exciton is much weaker for realistic values for the
lattice distortion.
The result for the N\'{e}el state is shown
in figure~\ref{heisfigextra}.
It is now seen that the $q=\pi/a$-exciton strongly dominates the
$q=0$-exciton. This is to be expected since the components
which imply a momentum transfer $q=0$ to the spin system
($Z_0$, $Z_M$) are much weaker now.
Even a large value of~$\delta$ does not drastically
change this situation.
It should be clear that 
the strengths of all exciton peaks above the 
exciton at~$\omega_{\pi/a}=U-V-W^2/(4V)$ 
are a direct measure for the dimerization
degree
{\em both\/} of the charge ($\delta,\eta \neq 0$) {\em and\/} the
spin ($\delta_S\neq 0$) system.

\section{Summary and conclusions}

In this paper we have further studied
the optical absorption for strongly-correlated electrons
in half-filled Peierls-distorted chains.
In (Gebhard {\em et al.} II 1996)
we exactly solved the case of the Harris-Lange model
with its incoherent spin background (``hot-spin case''). 
For the more realistic situation of low temperatures, i.~e.,
a unique ground state, exact solutions are impossible for the generic cases.
To make further progress we assumed that
the spin dynamics is irrelevant: (i)~we neglected corrections of
the order~$J$ to the optical 
excitation energies and thus a singlet/triplet-splitting
of the excitations, and (ii)~we assumed that no spin excitations are created
during an optical excitation (``no-recoil approximation'').
Then we could analytically investigate the problem even in the presence
of a lattice dimerization~$\delta$ and a nearest-neighbor interaction~$V$
between the electrons.

In the ``no-recoil approximation'' for the strongly correlated
Hubbard model all the features already seen in the ``hot-spin case''
become more prominent since now only transitions 
with $\Delta q_S=0$ and $\Delta q_S=\pi/a$ are allowed.
This corresponds to vertical transitions
between parallel Hubbard bands ($\Delta q_S=\Delta q_C=0$) and 
antiparallel bands ($\Delta q_S=\Delta q_C=\pi/a$).
The relative strength of the two processes can be measured
by three parameters ($Z_0$, $Z_{\pi}$, $Z_M$) which depend
on both the spin structure of the ground state and the lattice distortion.
In the generic case the parallel Hubbard bands result in 
a single peak at $\omega=U$.
The $\Delta q_S=\Delta q_C=\pi/a$ part supplements the optical
absorption spectrum by two absorption bands for
$W\delta \leq |\omega-U| \leq W$.
An additional nearest-neighbor interaction results
in a tightly bound exciton at $\omega_0=U-V$ and another exciton
at $\omega_{\pi/a}= U-V - W^2/(4V)$. 
Simpson's exciton band has seized to exist
since the exciton's center of mass momentum has to be 
$\Delta q_C=0$ or $\Delta q_C=\pi/a$.
Again, the excitons draw the oscillator strength from the ``band absorption''
part.
Although the location of the exciton resonances is independent
of the spin configuration of the ground state their
relative amplitude is a very sensitive measure of the lattice
and spin dimerization. We thus come to the important conclusion
that the linear {\em optical\/} absorption in strongly correlated
electron systems might serve as a subtle probe for the {\em magnetic\/}
structure of the ground state. In particular, optical absorption
allows to measure the ``hidden'' long-range order
of nearest-neighbor spin singlets (Talstra, Strong, and Anderson 1995). 

One might think that it is fairly simple
to experimentally distinguish between a  Peierls and a Mott-Hubbard insulator
on the basis of their optical absorption.
At low temperatures we expect a Peierls insulator to show
a broad band-to-band transition while
a Mott-Hubbard insulator with appreciable values for
the Coulomb interaction ($U \gg W$, $V >W/2$) should display
sharp exciton lines in the optical absorption spectrum.
Unfortunately, a direct comparison to experiment is
difficult for two reasons. First, disorder effects can inhomogeneously
broaden single lines. Hence, an experimentally
observed band can very well be a sign of disorder rather than
an argument against a Mott-Hubbard insulator.
Secondly, when a residual electron-electron interaction
is included in a Peierls insulator
(Abe, Yu, and Su 1992); (Abe, Schreiber, Su, and Yu 1992)       
one can equally well obtain excitons which draw the oscillator
strength form the band transitions. Hence, single exciton
lines are not a clear-cut indication against the presence
of a Peierls insulator either.

It is clear that one needs to compare model calculations and
experimental observations on many more physical
quantities than just the linear optical absorption
before definite conclusions can be reached.

\section*{Acknowledgments}

We thank H.~B\"{a}\ss ler, A.~Horv\'{a}th,
M.~Lindberg, S.~Mazumdar, M.~Schott, and
G.~Weiser for useful discussions.
The project was supported in part by the
Sonderforschungsbereich~383 
``Unordnung in Festk\"{o}rpern
auf mesoskopischen Skalen'' of the Deutsche Forschungsgemeinschaft.

\newpage
\begin{appendix}
\section*{Sum rule}
\setcounter{section}{1}

We briefly account for the sum rule. We have
\begin{equation}
\int_{0}^{\infty} d\omega\ {\rm Im}\{\chi(\omega)\}
= \pi \frac{{\cal N}_{\perp}}{La} \sum_n \left| \langle 0 |
\hat{\jmath}^2|n\rangle\right|^2 = \pi \frac{{\cal N}_{\perp}}{La}
\langle 0 | \hat{\jmath}^2|0\rangle \; .
\end{equation}
It is a standard exercise to show that
\begin{equation}
\langle 0 | \hat{\jmath}^2 | 0 \rangle =
(2tea)^2 \sum_l \left(1+(-1)^l\delta\right)^2\left(1+(-1)^l\eta\right)^2
\langle 0 | \left( \frac{1}{4}
-\hat{\rm\bf S}_l\hat{\rm\bf S}_{l+1}\right) | 0 \rangle \; .
\end{equation}
We define the positive quantities
\begin{equation}
C_S^{\rm even, odd}= \frac{1}{L} \sum_l \frac{1\pm (-1)^l}{2}
\langle 0 | \left( \frac{1}{4}
-\hat{\rm\bf S}_l\hat{\rm\bf S}_{l+1}\right) | 0 \rangle
\label{CSevenCSodd}
\end{equation}
such that $C_S=C_S^{\rm odd}+C_S^{\rm odd}$ is the 
value of the nearest-neighbor spin-spin correlation function.
Then we may write
\begin{equation}
\int_{0}^{\infty} d\omega\ {\rm Im}\{\chi(\omega)\}
= \pi {\cal N}_{\perp}a (2te)^2 \left[
\left(1+\delta\right)^2\left(1+\eta\right)^2
C_S^{\rm even}
+
\left(1-\delta\right)^2\left(1-\eta\right)^2
C_S^{\rm odd}
\right] \; .
\end{equation}
The area under the 
curves for~$\sigma_{\rm red}(\omega)$, equation~(\ref{sigmared}), 
is thus given by
\begin{equation}
\int_0^{\infty} \frac{d\omega}{W} \sigma_{\rm red}(\omega) =
\frac{\pi}{4}
\left[
\left(1+\delta\right)^2\left(1+\eta\right)^2 C_S^{\rm even}
+
\left(1-\delta\right)^2\left(1-\eta\right)^2 C_S^{\rm odd}
\right]
\label{sumrulesigmared}
\; .
\end{equation}
\end{appendix}

\newpage
\begin{center} {\bf REFERENCES} \end{center}

\begin{itemize}
\item S.~Abe, M.~Schreiber, W.~P.~Su, and J.~Yu,
Phys.~Rev.~B~{\bf 45}, 9432 (1992).
\item S.~Abe, J.~Yu, and W.~P.~Su, Phys.~Rev.~B~{\bf 45}, 8264 (1992).
\item L.~Alc\'{a}cer, in:
{\sl Organic Conductors}, ed. by J.-P.\ Farges, 
(Marcel Dekker, New York (1994)).
\item D.~Baeriswyl, P.~Horsch, and K.~Maki,
Phys.~Rev.~Lett.~{\bf 60} (C), 70 (1988).
\item D.~Baeriswyl, D.~K.~Campbell, and S.~Mazumdar, in: 
{\sl Conjugated Conducting Polymers}, ed.~by H.~Kiess,
(Springer Series in Solid State Sciences~{\bf 102}, Springer, Berlin (1992)).
\item A.~Brau and J.-P.~Farges, in:
{\sl Organic Conductors}, ed. by J.-P.\ Farges, 
(Marcel Dekker, New York (1994)).
\item D.~K.~Campbell, J.~T.~Gammel, and E.~Y.~Loh,
Phys.~Rev.~B~{\bf 38}, 12043 (1988).
\item D.~K.~Campbell, J.~T.~Gammel, and E.~Y.~Loh,
Int.~J.~Mod.~Phys.~B~{\bf 3}, 2131 (1989).
\item D.~K.~Campbell, J.~T.~Gammel, and E.~Y.~Loh, in: {\sl Interacting 
Electrons in Reduced Dimensions}, ed.~by D.~Baeriswyl and D.~K.~Campbell,
(NATO ASI Series~B~{\bf 213}, Plenum Press, New York (1989)), p.~171.
\item D.~K.~Campbell, J.~T.~Gammel, and E.~Y.~Loh,
Phys.~Rev.~B~{\bf 42}, 475 (1990).
\item E.~Dagotto, Rev.~Mod.~Phys.~{\bf 66}, 763 (1994).
\item P.~G.~J.~van Dongen, Phys.~Rev.~B~{\bf 49}, 7904 (1994).
\item P.~G.~J.~van Dongen, Phys.~Rev.~B~{\bf 50}, 14016 (1994).
\item F.~H.~L.~E\ss ler and V.~E.~Korepin (ed.), {\sl Exactly Solvable 
Models of Strongly Correlated Electrons}, (World Scientific, Singapore (1994)).
\item J.-P.\ Farges (ed.), {\sl Organic Conductors},
(Marcel Dekker, New York (1994)).
\item J.~L.~Fave, in: {\sl Electronic Properties of Polymers}, 
ed.~by H.~Kuzmany, M.~Mehring, and S.~Roth,
(Springer Series in Solid State Sciences~{\bf 107}, Springer, Berlin (1992)).
\item A.~Fritsch and L.~Ducasse, J.~Physique~I~{\bf 1}, 855 (1991).
\item R.~M.~Fye, M.~J.~Martins, D.~J.~Scalapino, J.~Wagner, and W.~Hanke,
Phys.~Rev.~B~{\bf 45}, 7311 (1992).
\item J.-P.~Galinar, J.~Phys.~C~{\bf 12}, L335 (1979).
\item J.~T.~Gammel and D.~K.~Campbell, Phys.~Rev.~Lett.~{\bf 60} (C), 
71 (1988).
\item F.~Gebhard, K. Bott, M. Scheidler, P.~Thomas, and S.~W.~Koch,
next to last article, referred to as~I.
\item F.~Gebhard, K. Bott, M. Scheidler, P.~Thomas, and S.~W.~Koch,
last article, referred to as~II.
\item A.~B.~Harris and R.~V.~Lange, Phys.~Rev.~{\bf 157}, 295 (1967).
\item H.~Haug and S.~W.~Koch, {\sl Quantum Theory of the Optical and 
Electronic Properties of Semiconductors}, (World Scientific, Singapore (1990)).
\item A.~J.~Heeger, S.~Kivelson, J.~R.~Schrieffer,
and W.-P.~Su, Rev.\ Mod.\ Phys.~{\bf 60}, 781 (1988).
\item J.~Hubbard, Proc.~R.~Soc. London, Ser.~A~{\bf 276}, 238 (1963).
\item S.~Kivelson, W.-P.~Su, J.~R.~Schrieffer, and A.~J.~Heeger,
Phys.~Rev.~Lett.~{\bf 58}, 1899 (1987). 
\item S.~Kivelson, W.-P.~Su, J.~R.~Schrieffer, and A.~J.~Heeger,
Phys.~Rev.~Lett.~{\bf 60} (C), 72 (1988).
\item W.~Kohn, Phys.~Rev.~{\bf 133}, A171 (1964).
\item S.~K.~Lyo and J.-P.~Galinar, J.~Phys.~C~{\bf 10}, 1693 (1977).
\item S.~K.~Lyo, Phys.~Rev.~B~{\bf 18}, 1854 (1978).
\item G.~D.~Mahan, {\sl Many-Particle Physics}, 
(2nd~edition, Plenum Press, New York (1990)).
\item P.~F.~Maldague, Phys.~Rev.~B~{\bf 16}, 2437 (1977).
\item S.~Mazumdar and S.~N.~Dixit, Phys.~Rev.~B~{\bf 34}, 3683 (1986).
\item F.~Mila, Phys.~Rev.~B~{\bf 52}, 4788 (1995).
\item M.~Ogata and H.~Shiba, Phys.~Rev.~B~{\bf 41}, 2326 (1990).
\item A.~Painelli and A.~Girlando, Synth.~Met.~{\bf 27},
A15 (1988).
\item A.~Painelli and A.~Girlando, in: {\sl Interacting Electrons in 
Reduced Dimensions}, ed.~by D.~Baeriswyl and D.~K.~Campbell, (NATO ASI 
Series~B~{\bf 213}, Plenum Press, New York (1989)), p.~165.
\item A.~Painelli and A.~Girlando, Phys.~Rev.~B{\bf 39}, 2830 (1989).
\item A.~Parola and S.~Sorella, Phys.~Rev.~Lett.~{\bf 60}, 1831 (1990).
\item L.~Salem, {\sl Molecular Orbital Theory of Conjugated
Systems}, (Benjamin, London (1966)).
\item M.~Schott and M.~Nechtschein, in: {\sl Organic Conductors},
ed. by J.-P.\ Farges, (Marcel Dekker, New York (1994)).
\item B.~S.~Shastry and B.~Sutherland, Phys.\ Rev.\ Lett.~{\bf 65}, 243 (1990).
\item W.~T.~Simpson, J.~Am.~Chem.~Soc.~{\bf 73}, 5363 (1951).
\item W.~T.~Simpson, J.~Am.~Chem.~Soc.~{\bf 77}, 6164 (1955).
\item C.~A.~Stafford and A.~J.~Millis, Phys.~Rev.~B~{\bf 48}, 1409 (1993).
\item C.~A.~Stafford, A.~J.~Millis, and B.~S.~Shastry, 
Phys.~Rev.~B~{\bf 43}, 13660 (1991).
\item J.~C.~Talstra, S.~P.~Strong, and P.~W.~Anderson
Phys.~Rev.~Lett.~{\bf 74}, 5256 (1995).
\item C.-Q.~Wu, X.~Sun, and K.~Nasu, Phys.~Rev.~Lett.~{\bf 59}, 831 (1987).
\item F.~C.~Zhang and T.~M.~Rice, Phys.~Rev.~B~{\bf 37}, 3759 (1988).

\end{itemize}

\typeout{figures}

\begin{figure}[th]
\caption{Reduced optical conductivity,
$\sigma_{\rm red}(\omega >0, V)$,
for the dimer state for $U=2W$ in the presence of a nearest-neighbor interaction,
$V=0,W/2,W$.
A broadening of~$\gamma=0.01W$ has been included.}
\label{heisfig01}
\end{figure}

\begin{figure}[th]
\caption{Reduced optical conductivity,
$\sigma_{\rm red}(\omega >0, V, \delta,\eta)$,
for the dimer state for $U=2W$
in the presence of a lattice distortion, $\delta=0.2$, $\eta=-0.06$, and 
a nearest-neighbor interaction, $V=0, W/2, W$.
A broadening of~$\gamma=0.01W$ has been included.}
\label{heisfig11}
\end{figure}

\begin{figure}[th]
\caption{Reduced optical conductivity,
$\sigma_{\rm red}(\omega >0, V, \delta,\eta)$,
for the N\'{e}el state for $U=2W$
in the presence of a lattice distortion, $\delta=0.2$, $\eta=-0.06$, and 
a nearest-neighbor interaction, $V=0, W/2, W$.
A broadening of~$\gamma=0.01W$ has been included.}
\label{heisfigextra}
\end{figure}


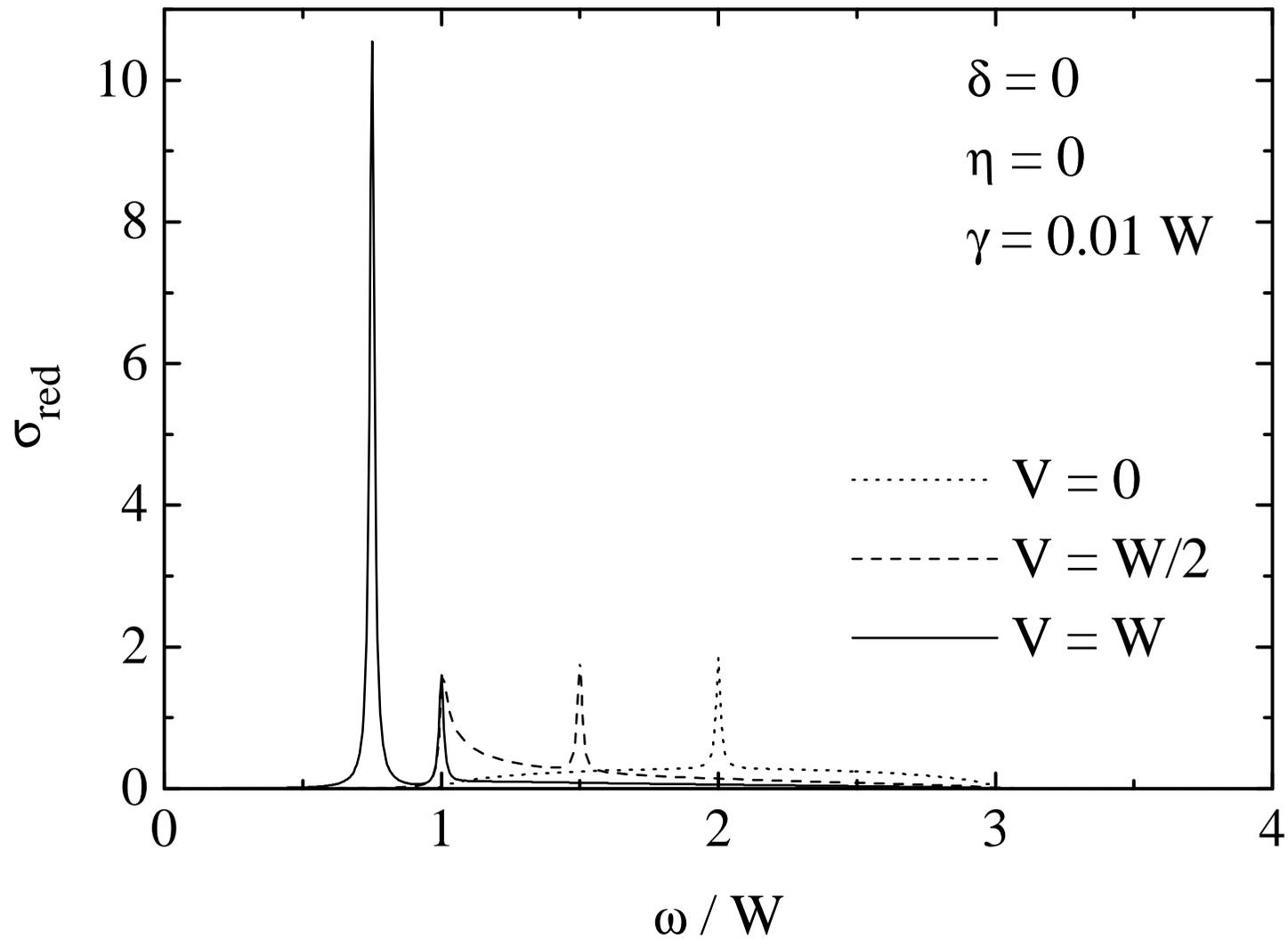
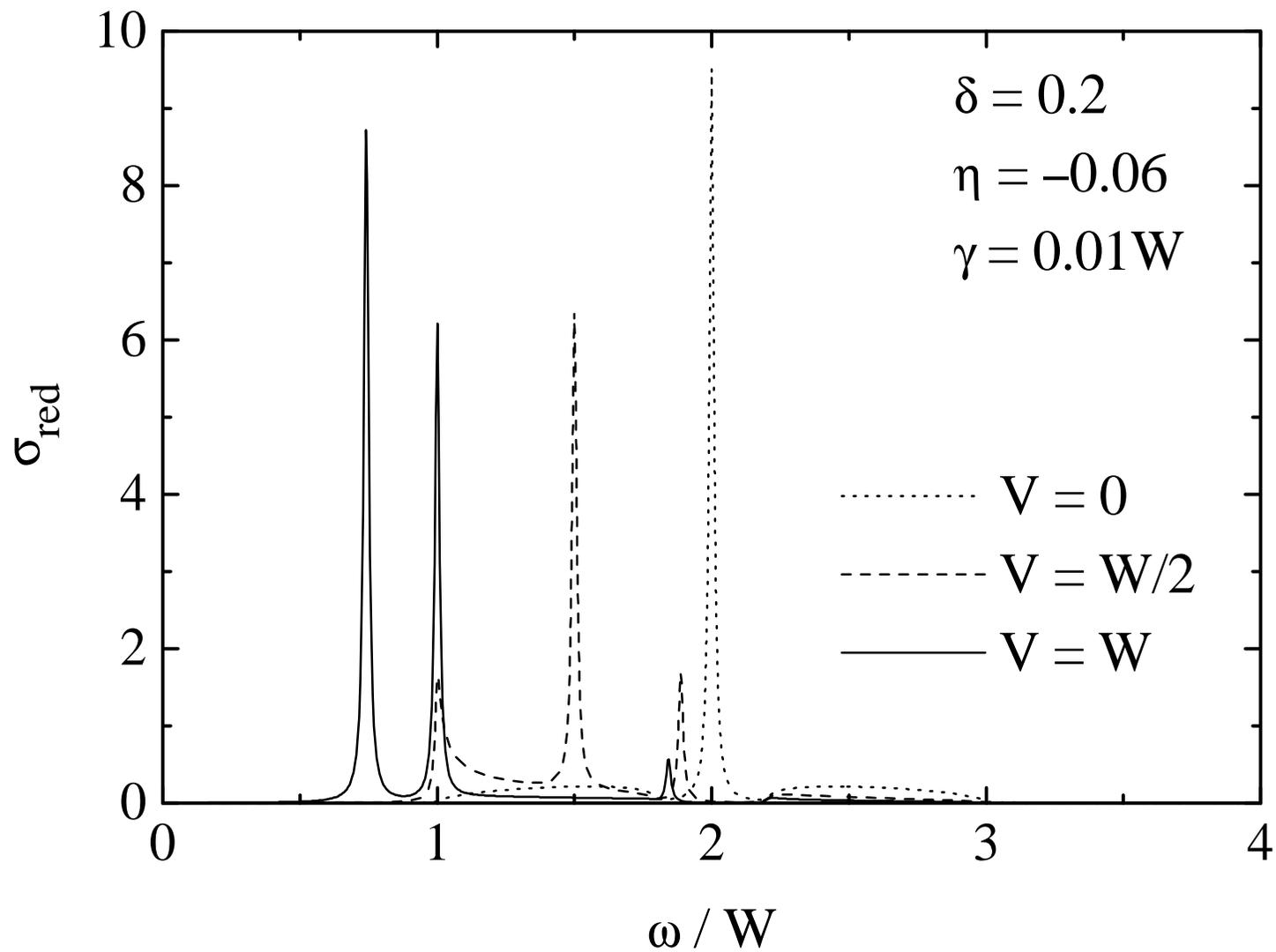
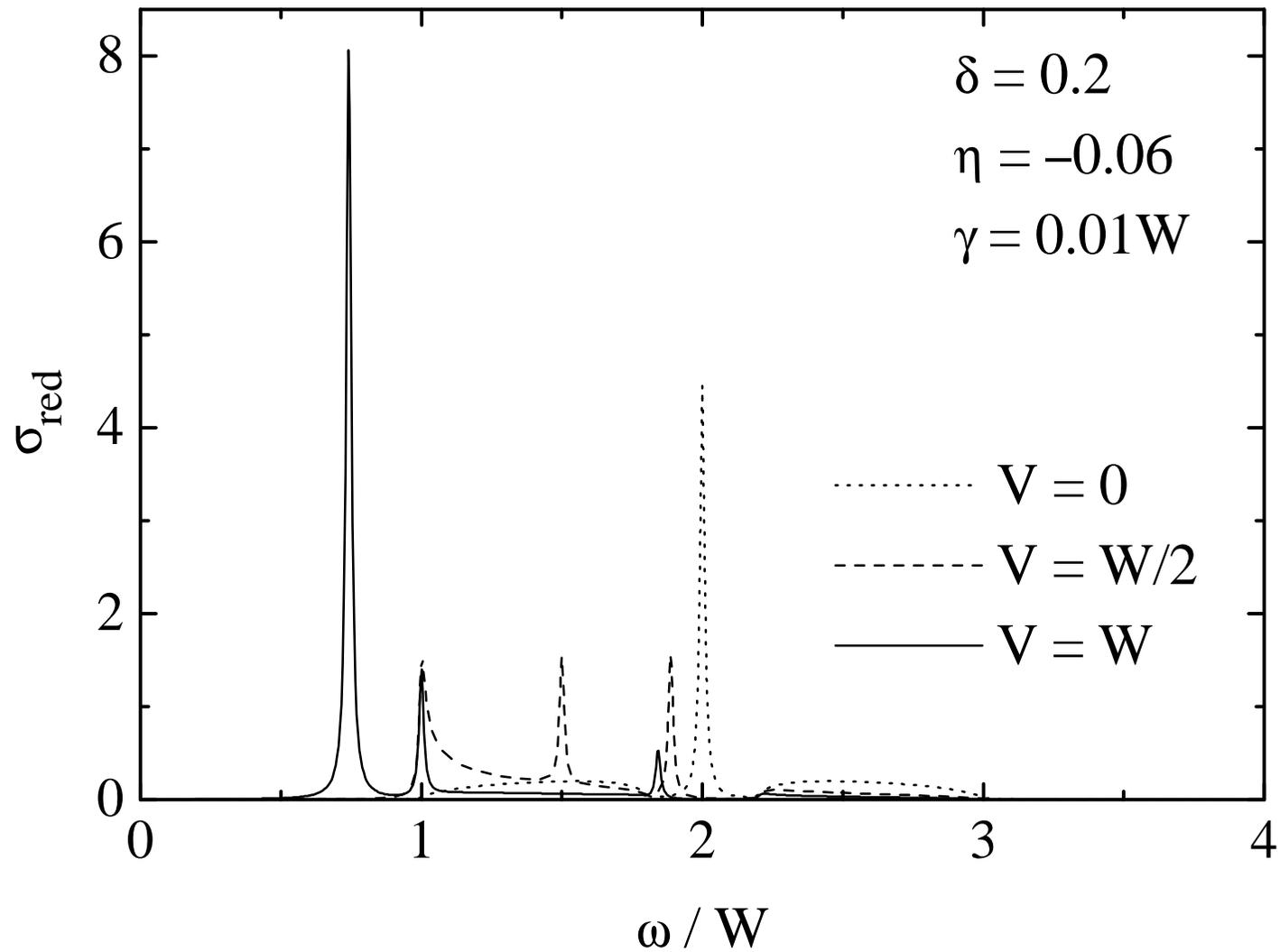
\end{document}